\documentclass[
reprint,
showpacs,
aps,amsmath,amssymb,
prl
]{revtex4-1}

\usepackage{graphicx}
\usepackage{dcolumn}
\usepackage{bm}

\usepackage{braket}
\def\ra{\rangle}
\def\la{\langle}
\def\bcen{\begin{center}}
\def\ecen{\end{center}}
\usepackage{epstopdf}
\begin{document}


\title{Duality and quantum state engineering in cavity arrays}

\author{Nilakantha Meher}
\author{S. Sivakumar}%
 \email{siva@igcar.gov.in}
\affiliation{%
 Materials Physics Division, Indira Gandhi Centre For Atomic Research,
  Homi Bhabha
 National Institute, Kalpakkam 603102,Tamilnadu,  India 
}%
\author{Prasanta K. Panigrahi}
 \email{pprasanta@iiserkol.ac.in}
\affiliation{%
Indian Institute of Science Education and Research Kolkata, Mohanpur 741246,
 West Bengal, India
}%


\date{\today}

\begin{abstract}
A system of two coupled cavities with $N-1$ photons is shown to be dynamically 
equivalent to an array  of $N$ coupled cavities containing one photon.  Every 
transition in the two cavity system has a dual  phenomenon in terms of photon 
transport in the cavity array.  This duality is employed to arrive at the 
required coupling strengths and nonlinearities in the cavity array so that 
controlled photon transfer is possible between any two cavities. This  transfer 
of photons between two of the cavities in the array is effected without 
populating the other cavities. The condition for perfect transport enables 
perfect state transfer between any two cavities in the array.    Further, 
possibility of high fidelity generation of generalized NOON states in two 
coupled cavities, which are dual to the Bell states of the photon in the cavity 
array, is established.

\end{abstract}

\maketitle


\section{Introduction}
 
Quantum theory provides for fundamentally newer ways of realizing secure 
communication, faster computation and precision metrology.  Quantum networks 
are basic to implementing these ideas.  Physical systems such as spin chains, 
cavity arrays, Joesphson junction arrays, quantum dots,  etc.,   have been 
investigated to explore their potential for network implementation.\\
Coupled-cavity arrays  have been used extensively in generating nonclassical 
states of the electromagnetic field and quantum information processing 
\cite{Har, Not}.  The technology has matured to such an extent that precise 
control of the cavity field dynamics is possible .   Suitable tailoring of 
inter-cavity coupling has been put to use in  entanglement generation  \cite{Miry},
 quantum state preparation \cite{Lar,Almeida}, state transfer between spatially 
 separated  cavities \cite{Cir, Verm}, exhibition of quantum interference 
 \cite{Nil, Nil2}, Rabi oscillation \cite{Yoshi}, to mention a few.   
 Consideration of cavities filled with nonlinear media has led to exploration of 
other quantum phenomena \cite{DEC},  such as  photon blockade \cite{Ima, Birn, Ada}
 and localization-delocalization related to bunching and antiubunching of photons 
\cite{Nil,Hen,Sch}. Interplay between intra-cavity nonlinearity and inter-cavity
coupling strength has been exploited to control the photon  statistics in cavity 
\cite{Sara}.  Engineering of inter-cavity coupling in arrays offers the possibility 
of simulating  effects of disorder \cite{Seg}, phase transitions 
\cite{Hart, Hart2, Bran},  etc. in condensed matter physics.\\
The identical and indistinguishable nature of photons require that the number 
of photons and the number of levels to be occcupied by them are considered 
together. For example, on the consideration of blackbody radiation, Planck 
distribution is obtained for $N$ photons to be distributed in $g$ levels when 
the number of possible ways of distributing as $(N+g-1)!/N!(g-1)!$. It is 
interesting to note that the result is the same if there are $g-1$ photons 
and $N+1$ levels.  This possibility of interchanging the roles of the number 
of particles and the number of levels is a duality.  Another simple example of 
duality is in  Euler characteristic $V-E+F=2$, where $V,E$ and $F$ refer to the 
number of vertices, edges and faces respectively of a convex solid.  In this 
expression the roles  of $V$ and $F$ are interchangeable.  Such duality 
relationships are very much sought after in physics, which often facilitates 
understanding of nontrivial aspects of one system in terms of easily accessible 
features of the other \cite{atiyah}.    In this article, a duality is 
established between two dynamical systems, namely,  one photon in an array of 
$N$ cavities and $N-1$ photons in two coupled cavities,  considering both linear 
and Kerr-nonlinear cavities.  Every transition in the two cavity system has a 
dual  phenomenon in terms of photon transport in the cavity array.   This feature 
helps to identify the conditions required in the linear cavity array  for a  
perfect transport  of photon between two cavities equidistant from the respective 
ends.  The result is generalized to nonlinear cavities which allows perfect 
transport between  any two cavities in the array.    Another prospect that makes 
this study  interesting in the context of information transfer is the possibility 
of perfect state transfer from one cavity to another.  Our results also point 
to the possibility of generating NOON states, which have found diverse 
applications and relate to a Bell state of the dual system 
\cite{Dor, Keb, Sean, Marcel}.

\begin{figure}
\centering
\includegraphics[scale=.25]{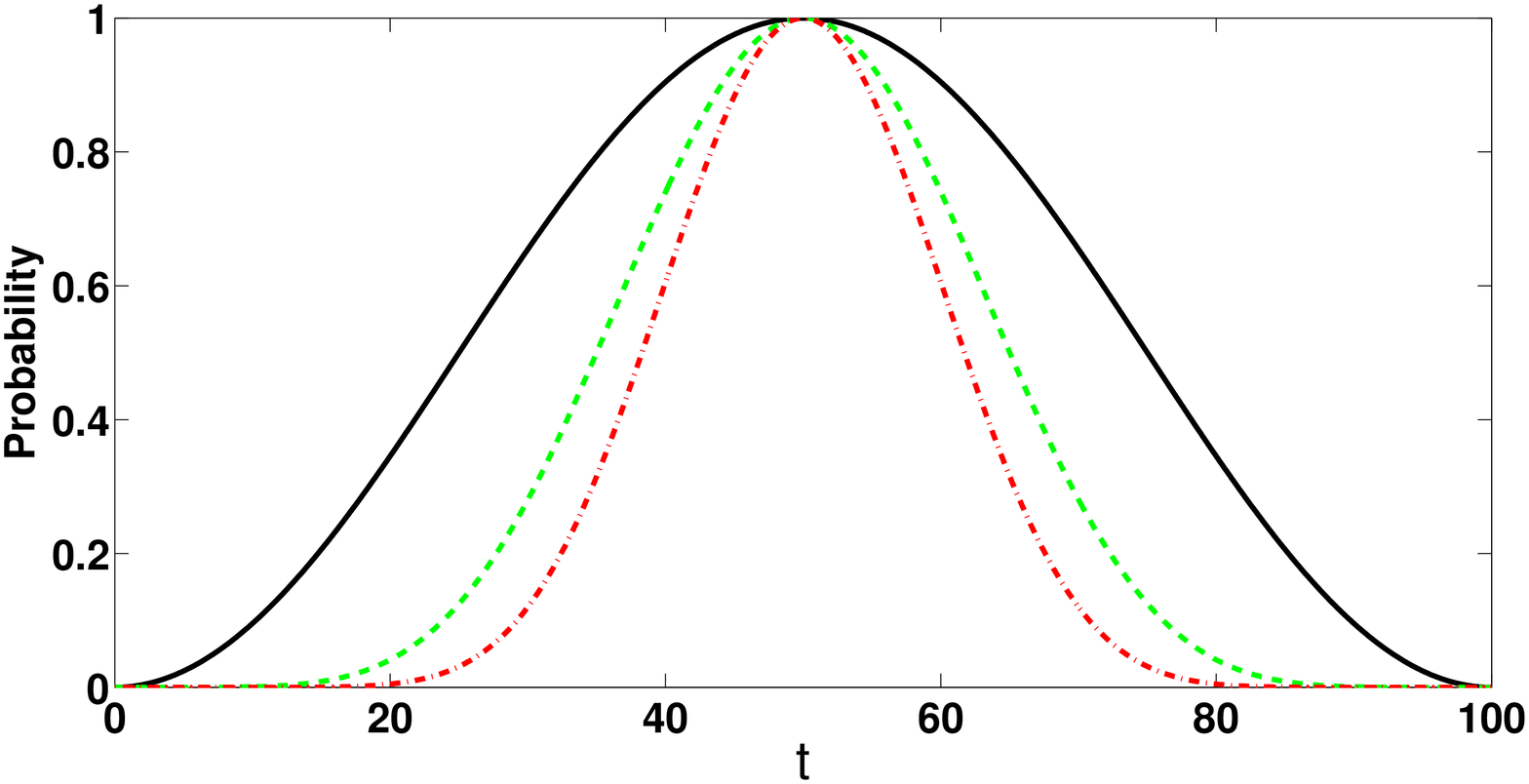}
\caption{Time evolution of probability for the coupled cavity to be 
in $\ket{0,N-1}$ on evolution from the initial state $\ket{N-1,0}$, 
with  $J=10^{-2}\pi$.  By duality, these profiles  show the probability
 of detecting a single photon in $N$th (end) cavity in the cavity array.
   Different curves correspond to  
   $N=2$ (continuous), 4 (dashed)and  6 (dot-dashed).}
\label{Endstatecavities}
\end{figure}

\begin{figure}
\centering
\includegraphics[scale=.25]{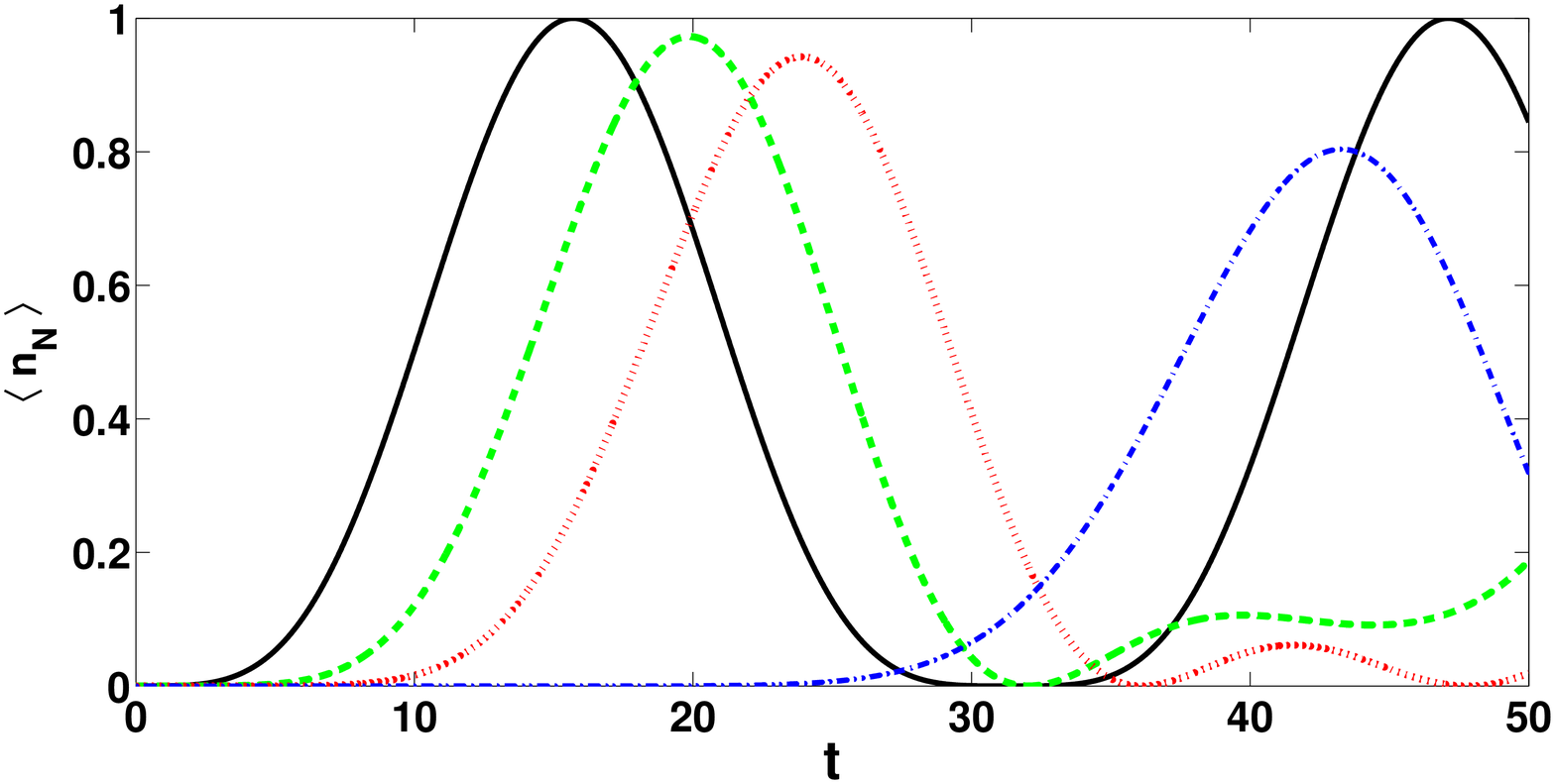}
\caption{Average number of photon in the end cavity as a function
 of t in cavity array. Number of cavities in the array is 
 $N=$3 (solid line), 4 (dashed), 5 (dotted) and 10 (dot-dashed). 
 All the cavities are identical and homogeneously coupled with 
 coupling strength $J=\sqrt{2}/10$. The system parameters are 
 normalized in the unit of $\tilde{\omega}$.}
\label{uniformCoupling}
\end{figure}

Consider a system of two linearly coupled cavities described by the Hamiltonian  
\begin{equation}\label{HJC}
H_{A}=\omega_1 a_1^{\dagger}a_1+\omega_2 a_2^{\dagger}a_2+
J\left[a_1^{\dagger}a_2+a_1a_2^{\dagger}\right]. 
\end{equation}
Here $\omega_1$ and $\omega_2$ are the resonance frequencies of the respective
 cavities  and $J$ is the coupling strength.  Suffix $A$ has been used to refer
  to this system of two coupled cavities.
Here $a_{1(2)}$ and $a_{1(2)}^{\dagger}$ are the annihilation and creation 
operators for the first(respectively, second)  cavity.  
Let $\vert n+1\ra$ represent the bipartite state $\vert N-1-n,n\ra$ of the two 
cavities corresponding to $N-n-1$ quanta in the first cavity and $n$ quanta 
in the second cavity. The total number of quanta in the two cavities is $N-1$. 
If the number of photons is fixed to be $N-1$, the Hamiltonian is expressed as
\begin{align}
H_A=&\sum_{n=0}^{N-1}\Omega_{n+1}\ket{n+1}\bra{n+1}\nonumber\\
&+\sum_{n=0}^{N-2}J_{n+1}(\ket{n+1}\bra{n+2}
+\ket{n+2}\bra{n+1}),
\end{align}
where $\Omega_{n+1}=[(N-1-n)\omega_1+n\omega_2]$ and 
$J_{n+1}=\sqrt{(n+1)(N-1-n)}J$.  \\
Now consider a system of $N$ linearly coupled cavities, described by  
\begin{equation}\label{cavityarrays}
H_{B}=\sum_{l=1}^N \tilde{\omega}_l b_l^\dagger b_l +\sum_{l=1}^{N-1} 
\tilde{J}_l(b_l^\dagger b_{l+1}+b_l b_{l+1}^\dagger),
\end{equation}
where $\tilde{\omega}_l$ is the cavity resonance frequency for the $l$th cavity, 
$b_l$ and $b_l^{\dagger}$ the  annihilation and creation operators for 
the $l$th cavity.  The strength of coupling between the $l$ and $(l+1)$ 
cavities in the array is $\tilde{J}_l$.  This form of the Hamitonian can 
be mapped to that of a spin network which has been studied in the context 
of state transfer and entanglement generation \cite{bose}.
For a single quantum in the system, the possible states are $\vert l\ra\ra$, 
which represents one photon in the $l$th cavity while the other cavities are
 in their respective vacuua.  Then the Hamiltonian is  
\begin{align}
H_{B}=\sum_{l=1}^N\tilde{\omega_l}\ket{l}\ra\la\bra{l}+
\sum_{l=1}^{N-1}\tilde{J}_l(\ket{l}\ra\la\bra{l+1}+\ket{l+1}\ra\la\bra{l}).
\end{align}
Duality of the two systems described by $H_A$ and $H_B$ respectively is 
identified if  $\tilde{J}_l=\sqrt{l(N-l)}J$, $\tilde{\omega_l}=
[(N-l)\omega_1+(l-1)\omega_2]$ and $l=n+1$. The transition 
$\ket{N-1-n,n}\rightarrow\ket{N-2-n,n+1}$ in the system of two 
cavities corresponds to photon transport from 
$\vert n+1\ra\ra\rightarrow\vert n+2\ra\ra$ in the array. In essence, 
transitions in the two-cavity system are equivalent to transport of a 
photon across the cavities in the array.    
\\
If the initial state of the two cavity system at resonance 
$(\Delta=\omega_1-\omega_2=0)$ is $\ket{N-1-n,n}$, it evolves to 
\begin{align}\label{UnitEvol}
e^{-iH_At}\ket{N-1-n,n}=&e^{-i(N-1)\omega t}\sum_{k=0}^{N-1-n} 
\sum_{l=0}^{n}{^{N-1-n}C_k}{^nC_l}\nonumber
\\
&(\cos Jt)^{N-1-(k+l)}(-i\sin Jt)^{k+l}\nonumber\\
 \sqrt{\frac{^{(N-1)}C_n}{^{(N-1)}C_{n+k-l}}}&\ket{N-1-(n+k-l),n+k-l},
\end{align}
at time $t$. It is worth noting that the time-evolved state is an atomic 
coherent state \cite{Kim}. \\
At $t=\pi/2J$ the time-evolved  state is $\ket{n,N-1-n}$, corresponding to 
swapping the number of photons in the cavities.  Time evolution of the 
respective probabilities for $\ket{N-1,0}$ to become $\ket{0,N-1}$ 
corresponding to $N=2,4$ and $6$ are shown in Fig. \ref{Endstatecavities}.
 Complete transfer of photons  between the end cavities of the array  
 corresponds to $\vert N-1,0\ra\rightarrow\vert 0,N-1\ra$ transition in 
 the coupled cavity system. By the duality between $H_A$ and $H_B$, these 
 profiles also represent the probability of transferring a photon from one end 
 to the other in an array of $2,4$ and $6$ cavities respectively. It may be 
 noted that the probabilities attain their peak value of unity, corresponding
  to complete transport of a quantum between the end cavities, when  
  $t=\pi/2J$.\\
It is essential that $J_l$ are related to $J$ in the specified manner.  
It is of interest to note that the requirement for such inhomegeneous 
couplings in linear quantum spin networks, optical waveguide arrays has 
been explored \cite{Chris,Yogesh}. If the coupling strengths $J_l$ 
are equal and all the cavities are identical, the average number of quanta at time $t$ in the $j$-th cavity is given by
 \begin{equation}
\langle n_j\rangle_t=\langle b_j^\dagger b_j\rangle_t=
\sum_{l=1}^N |G_{jl}|^2 \langle b_l^\dagger b_l\rangle_0.
\end{equation}
where,
\begin{align}
G_{jl}=&\sqrt{\left(\frac{2}{N+1}\right)}
\sum_{k=1}^N e^{-i(\tilde{\omega}+2J\cos(\frac{\pi k}{N+1}))t}\nonumber\\
&\sin\left(\frac{j\pi k}{N+1}\right)\sin\left(\frac{l\pi k}{N+1}\right).
\end{align}  
Here $\tilde{\omega}$ is the resonance frequency of the cavities in the array.
If the quantum is initially in the first cavity, that is,  
$\langle b_l^\dagger b_l\rangle_0=\delta_{1,l}$, then  
 \begin{equation}
\langle n_N\rangle_t=\langle b_N^\dagger b_N\rangle_t=|G_{N1}|^2,
\end{equation}
is the average photon number in the last cavity. For large $N$, 
$\sin(Nk\pi/N+1)\approx \sin(k\pi)=0$ and $G_{N1}$ tends to zero.
 This indicates that complete transfer is not possible. 
Time evolution of the average number of quantum $\langle n_N\rangle$
 in the end cavity for arrays with $N=$3, 4, 5 and 10 cavities 
 respectively are shown in Fig. \ref{uniformCoupling}.   From the figure,
 it is clear that complete transfer occurs if the array has  three cavities.
 Maximum of $|G_{N1}|^2$ decreases with increasing the number of cavities.
   Hence, complete transfer does not occur if the homegeneously coupled 
   array has more than three cavities whereas inhomogeneous coupling achieves
   complete transfer in shorter time \cite{Chris,Chris2}.\\   

 It is to be further noted that complete transition  is possible only between
  the states $\ket{N-1-n,n}$ and  $\ket{n,N-1-n}$ of the coupled cavities. 
  Analogously complete transfer of a single photon  can occur only in between 
  $(n+1)$th and $(N-n)$th cavities  in the cavity array. With linear coupling,
   it is not possible to achieve complete transfer between two arbitrary 
   cavities in the array. \\
     
To see if nonlinearity helps in steering the evolution of states to achieve
 perfect transfer, we consider the Kerr-type nonlinearity. We present the
 analysis of two coupled nonlinear cavities. The required Hamiltonian is  
\begin{equation}\label{Hamilt}
H'_A=\omega_1 a_1^{\dagger}a_1+\omega_2 a_2^{\dagger}a_2+\chi _1 
a_1^{\dagger 2} a_1^2+\chi_2 a_2^{\dagger 2}a_2^2+J\left[a_1^{\dagger}a_2+a_1a_2^{\dagger}\right],
\end{equation}
which describes two linearly coupled Kerr cavities.\\

If it is required to evolve from $\vert m,n\ra$ to $\vert p,q\ra$,  
consider the superposition $\ket{X_\pm}=1/\sqrt{2}(\ket{m,n}\pm\ket{p,q})$.   
These two states  become approximate eigenstates of the $H'_A$ if 
$J << \chi_1, \chi_2, \omega_1,\omega_2$, and
\begin{align}\label{OptCond}
\Delta=\frac{(p(p-1)-m(m-1))\chi_1+(q(q-1)-n(n-1))\chi_2}{m-p}.
\end{align}
This condition is equivalent to  
\begin{equation}\label{avgenergy}
\la m,n\vert H'_A\vert m, n\ra= \la p,q\vert H'_A\vert p, q\ra.
\end{equation}
This equality of average energy in the two states is another way of 
stating  the requirement that the states $\ket{X_\pm}$ are approximate 
eigenstates of $H'_A$.   In the discussion that follows it is assumed 
that $\chi_1=\chi_2=\chi>0$ and the condition simplifies to  $\Delta=2\chi(n-p)$. \\

     If the initial state is $\ket{m,n}=1/\sqrt{2}(\ket{X_+}+\ket{X_-})$,
      the state of the system at a later time is,
$\ket{\psi(t)}\approx\left[\cos\left(\theta_a t\right)\ket{m,n}
-i\sin\left(\theta_a t\right)\ket{p,q}\right],$
with $\theta_a=(\lambda_s^a-\lambda_n^a)/{2}$. Here $\lambda_s^a$ and 
$\lambda_n^a$ are the eigenvalues of $H'_A$ corresponding to the approximate 
eigenvectors $\vert X_+\ra$  and $\vert X_-\ra$ respectively.    
At $t=\pi/(\lambda_s^a-\lambda_n^a)$, the time-evolved state is $\vert p,q\ra$.  
This is the minimum time required to switch  from $\vert m,n\ra$ to 
$\vert p,q\ra$.  Thus, the state switching (SS) condition given in 
Eq. \ref{OptCond} ensures that there is complete transfer from the 
initial state $\vert m,n \ra$ to the desired final state $\vert p,q \ra $.  \\
\begin{figure}
\centering
\includegraphics[scale=.25]{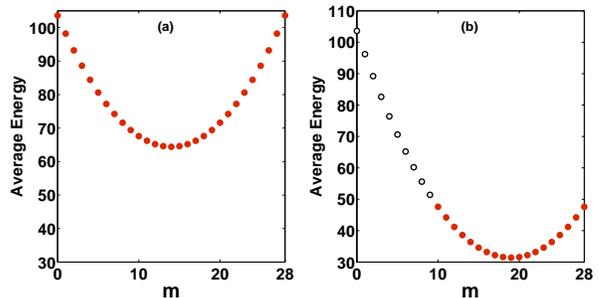}
\caption{Average energy $\bra{m,n} H'_A\ket{m,n}$ as a function
 of $m$, with  $\Delta=0,\chi=0.1$ (Left) and $\Delta=-2,\chi=0.1$
  (Right). The system parameters are expressed in units of $\omega_1$.}
\label{AvgPotential}
\end{figure}
It is immediate that detuning $\Delta$ and nonlinear coupling 
 strength $\chi$ can be properly chosen for a given value for $n-p$. 
  As the value of $n$ is specified in the initial state $\vert m,n\ra$, 
  the two parameters $\Delta$ and $\chi$ fix the number of quanta say $s$, 
 that can be transferred and the target state becomes $\ket{p=m\pm s,q=n\mp s}$. 
 It needs to be emphasized that for a given $\Delta$ and $\chi$ satisfying the 
 SS condition, no more than two states can have their  average energies equal 
 as shown in Fig. \ref{AvgPotential}. Once these parameters are fixed, 
 probability of transition to any state other than the target state is 
 negligible.  Hence, $\Delta$ and $\chi$ provide control to steer the 
 system from  the initial state $\vert m,n\ra$  to the final state 
 $\vert p,q \ra$.  \\

In Fig. \ref{AvgPotential}, $\bra{m,n}H'_A\ket{m,n}$ is plotted as a function
 of $m$, keeping $m+n=28$ fixed. From Fig. \ref{AvgPotential}(a),  it is seen
  that every state has only one partner state with equal average energy.  
  So, SS can occur between these partner states.  It is observed from 
  Fig. \ref{AvgPotential}(b) that not every state has a partner state 
  with equal average energy.    Essentially, states without partner states 
  are approximate eigenstates of $H'_A$ and, therefore, do not evolve.  
  This brings out another control aspect available in the system, namely, 
  the possibility of inhibiting evolution of certain states with properly 
  chosen values of the  control parameters $\chi$ and $\Delta$.

\begin{figure}
\centering
\includegraphics[scale=0.25]{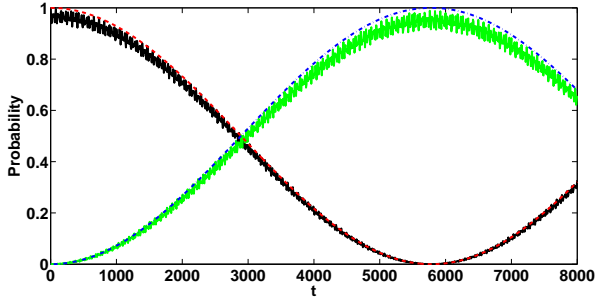}
\caption{Probability of detecting the state $\ket{14}$ from $\ket{50}$
 as a function of $t$. Detecting other states are practically zero. 
 Continuous black(dashed) and continuous green (dashed-dot) line 
 corresponds to $P_{50}$ and $P_{14}$ calculated numerically 
 (approximate  analytical solution $\ket{\psi(t)}$). We set
  $\Delta=-0.2$, $\chi=0.1$, $J=0.035$. The system parameters 
  are used in the unit of $\omega_1$.}
\label{QswitchP50P14}
\end{figure}

   Consider the initial state of the coupled cavity system to be $\ket{50}$.
  In Fig. \ref{QswitchP50P14}, the probability of detecting the system 
  in the state $\ket{14}$  at later times is shown when the required SS 
  condition is satisfied.  The values have been generated from the approximate
   evolved state $\ket{\psi(t)}$ and also by exact numerical solution of  the 
   evolution corresponding to $H'_A$.  It is seen that the quanta are exchanged
   periodically driving the system between  $\ket{14}$ and $\ket{50}$ and  
   transfer to other states is insignificant. \\

\begin{figure}
\centering
\includegraphics[scale=0.25]{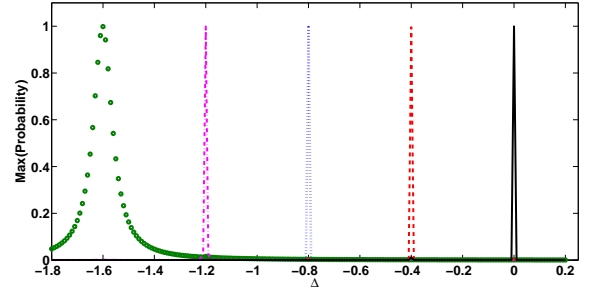}
\caption{Maximum probability of detecting quantum states $\ket{p,q}$
  as a function of $\Delta$ for $\chi=0.2$ and $J=0.05$ from the 
  initial state $\ket{5,0}$. Note that, complete switching occurs 
  from $\ket{5,0}$  to $\ket{41}$ (circle), $\ket{32}$ (dot-dashed),
   $\ket{23}$ (dotted), $\ket{14}$(dashed) and $\ket{05}$ (continuous
   ) at $\Delta=-8\chi$, $\Delta=-6\chi$, $\Delta=-4\chi$, 
   $\Delta=-2\chi$ and $\Delta=0$ respectively satisfying the 
   relation Eqn. \ref{OptCond}. The system parameters are expressed 
   in the unit of $\omega_1$.  }
\label{DelVsProb}
\end{figure}

In order to effect transition to other states, the value of $\Delta$ can be 
chosen properly.  The maximum probabilities of detecting the target state 
$\ket{p,q}$ with $p=1,2,3,4,5$ and $p+q=5$ from $\ket{50}$ are shown in 
Fig. \ref{DelVsProb}  as a function of $\Delta$. The value of $\chi$ has
 been chosen to be 0.2. Depending on the value of detuning, exchange of 
 quanta is precisely controlled to different target states. \\     

	A duality relation of the two cavity system with  the cavity array 
	system is possible in this nonlinear case too. Consider the 
	nonlinear cavity array Hamiltonian  
\begin{align}\label{Nonlinearcavityarrays}
H'_{B}=&\sum_{l=1}^N \tilde{\omega}_l b_l^\dagger b_l 
+\tilde{\chi}_l (b_l^{\dagger} b_l)^2 \nonumber \\
&+\sum_{l=1}^{N-1} \sqrt{l(N-l)}J(b_l^\dagger b_{l+1}+b_l b_{l+1}^\dagger),
\end{align}
which includes Kerr nonlinearity in each cavity of the  array.  This is dual
 to $H'_A$ if $\tilde{\omega}_{k+1}+\tilde{\chi}_{k+1}=
 (N-1-k)\omega_1+k\omega_2+[(N-1-k)(N-2-k)+k(k-1)]\chi$.   
With this identification, transitions among levels  in the two Kerr cavity 
system  can be mapped to transfer of photon in the Kerr cavity array. \\

In particular, transition from $\ket{N-1-n,n}$ to $\ket{N-1-q,q}$  in the 
coupled cavities  corresponds to transferring a  photon between 
$(n+1)$-cavity to $(q+1)$-cavity in the cavity array.   The condition to
 realize this transfer is 
 $\la\la n+1\vert H'_B\vert n+1\ra\ra=\la\la q+1\vert H'_B\vert q+1\ra\ra$,
  whose dual relation for the coupled cavities is given in Eq. \ref{avgenergy}.
    For the Hamiltonian $H'_B$, this condition yields,
\begin{align}\label{SScavityarrays}
\tilde{\chi}_{k+1}+\tilde{\omega}_{k+1}&
=(N-1)\omega_1-2k\chi(n+q+1-N)\nonumber\\
&+[(N-1-k)(N-2-k)+k(k-1)]\chi
\end{align}
 to realize complete transfer of photon occurs between the cavities.
  On employing cavity-dependent nonlinearity $\tilde{\chi}_l$, 
  controlled transfer of photons between selected cavities is achievable.
     Such site-dependent nonlinearity has been realized recently by embedding
      quantum dots in each photonic crystal cavities \cite{Calic,Hugg, Surr}. \\

In the limit of weak coupling strength $J$, 
$\frac{1}{\sqrt{2}}(\ket{n+1}\ra\pm\ket{q+1}\ra)$ are  eigenstates of $H'_B$
 and the corresponding eigenvalues are denoted by $\lambda_s^b$ and
  $\lambda_n^b$.  The initial state $\ket{n+1}\ra$ evolves under $H'_B$ to
$\ket{\psi(t)}\ra\approx\cos\left(\theta_b t\right)\ket{n+1}\ra 
-i\sin\left(\theta_b t\right)\ket{q+1}\ra,$
where $\theta_b=(\lambda_s^b-\lambda_n^b)/{2}$.
It is seen that the photon is exchanged periodically between the cavities.
   An important   feature of this process is that the other cavities in the array
     are not populated to any appreciable extent during the evolution.  
     This conclusion is based on the observation  that the states other 
     than $\ket{n+1}\ra$ and $\ket{q+1}\ra$ do not contribute appreciably 
     to $\ket{\psi (t)}\ra$. \\ 

 If the coupling term in the Hamiltonian $H'_A$ is 
taken to be $J\left[e^{i\eta}a_1^{\dagger}a_2+e^{-i\eta}a_1a_2^{\dagger}\right]$
 making the coupling constants are complex, then the initial state  $\ket{m,n}$, 
evolves to $\ket{\psi(t)}\approx\left[\cos\left(\theta_a t\right)\ket{m,n}
-ie^{-i(q-n)\eta}\sin\left(\theta_a t\right)\ket{p,q}\right].$ These states are  
of the form $\ket{\psi}=\cos\theta\ket{m,n}+e^{i\phi}\sin\theta\ket{p,q}$, if  
$\theta=\theta_a t$ and $\phi=-(\pi/2+(q-n)\eta)$. \\

If $m\ne p$ and $\theta\ne 0,\pm\pi/2$, then $\ket{\psi}$ is entangled.
  Additionally, if $m, q=N$ and $\theta=\pi/4$,  the resultant state is 
\begin{align*}
\ket{\psi}&=\frac{1}{\sqrt{2}}(\ket{N0}+e^{i\phi}\ket{0N}),
\end{align*}
the generalized NOON state. In the case of cavity array this is equivalent
 to generating the Bell state $\ket{\psi}\rangle=\cos\theta\ket{n+1}\ra+
 e^{i\phi}\sin\theta\ket{q+1}\ra$.\\

 Another important outcome of the complex coupling  in the context of single
  photon in cavity array is the possibility of state transfer between {\it any}
   two cavities.      Consider the initial state of the cavity array to be 
   $\alpha\vert\mbox{vac}\ra\ra+\beta\ket{n+1}\rangle$, which corresponds to 
   the $(n+1)$-th cavity in the superposition  
   $\alpha\vert 0\ra+\beta\vert 1\ra$ and the other cavities are 
   in their respective vacuua.   If the SS condition is satisfied,   the 
   time-evolved state is $\alpha\vert\mbox{vac}\ra\ra+\beta e^{-i\lambda t}
   (\cos\theta_bt\ket{n+1}\ra-ie^{-i\eta (q-n)}\sin\theta_bt\ket{q+1}\ra)$, 
   where $\lambda=(\lambda_s^b+\lambda_n^b)/2$.  At $2\theta_b t=\pi$, 
   then the state of the $(q+1)$-th cavity is  the superposition 
   $\alpha\vert 0\ra+\beta\vert 1\ra$ and the other cavities in their 
   respective vacuua for the suitable value of $\eta$.  Thus,
    the SS condition ensures the state of the field in the $(n+1)$-th
     cavity is transferred to the  $(q+1)$-th cavity.

To summarize, transfer of a photon in an array of $N$  cavities is dynamically
 equivalent or dual  to the problem of sharing of ($N-1$) photons between two 
 coupled cavities, provided the parameters relevant to the systems are chosen 
 properly. This duality is extendable even if the cavities are of Kerr-type, which,
 in turn, requires the couplings to be inhomogeneous.  Duality between 
 the two systems has made it transparent to identify the correct combination 
 of the coupling strengths and local nonlinearities in the array for complete
 photon transfer between any 
two cavities in the array.  In the linear case, perfect transport is possible 
only between the cavities which are symmetrically located from the end 
cavities of the array. Kerr nonliearity allows perfect transport between 
any two cavities without any restriction whatsoever. Importantly, this 
transfer is effected without populating the other cavities in the array, 
so that the transfer cannot be viewed as a continuous hopping of photons 
from one cavity to the other. In the presence of Kerr nonliearity and 
complex coupling  strengths among the cavities, perfect state transfer 
between any two cavities is achieved.  This feature is important in the 
context of encoding and transfer of information.    High fidelity 
generation of entangled states of the form  
$\cos\theta\ket{m,n}+e^{i\phi}\sin\theta\ket{p,q}$ in coupled 
cavities is possible in the presence of Kerr nonlinearity in the 
cavities.  Equivalently, Bell states of the cavity array are also 
achievable with high fidelity.   The results of this work are 
important for designing suitable experiments for controlled transfer
 to photons by quantum switching.\\

One of the authors (NM) acknowledges the Department of Atomic Energy of 
the Government of India for a senior research fellowship.




\begin{thebibliography}{40}
\bibitem{Har}{S. Haroche and J.-M. Raimond, \textit{Exploring 
the Quantum: Atoms, cavities, and Photons} 
(Oxford university Press, New York, 2006).}
\bibitem{Not}{M. Notomi, E. Kuramochi, and T. Tanabe, 
Nature Photonics \textbf{2}, 741 (2008).}
\bibitem{Miry}{ S. R. Miry, M. K. Tavassoly, R. Roknizadeh, 
 Quantum Inf. Process. \textbf{14}, 593 (2015).}
\bibitem{Lar}{J. Larson and E. Andersson, Phys. Rev. A {\bf 71}, 
053814 (2005).} 
\bibitem{Almeida}{G. M. A. Almeida, F. Ciccarello, T. J. G. Apollaro, 
and A. M. C. Souza, Phys. Rev. A {\bf 93}, 032310 (2016).}
\bibitem{Cir}{J. I. Cirac, P. Zoller, H. J. Kimble, and H. Mabuchi, 
Phys. Rev. Lett. {\bf 78}, 3221 (1997).}
\bibitem{Verm}{B. Vermersch, P.-O. Guimond, H. Pichler, and P. Zoller
 Phys. Rev. Lett. \textbf{118}, 133601 (2017).}
\bibitem{Nil}{N. Meher and S. Sivakumar, J. Opt. Soc. Am. B 
\textbf{33}, 1233 (2016). }
\bibitem{Nil2}{N. Meher and S. Sivakumar, 13th International Conference 
on Fiber Optics and Photonics, OSA Technical Digest (online) 
(Optical Society of America, 2016), paper Th3A.73.}
\bibitem{Yoshi}{Y. Sato, Y. Tanaka,	J. Upham, Y. Takahashi,	T. Asano,
  S. Noda, Nature Photonics, \textbf{6}, 56 (2012).}
\bibitem{DEC}{D. E. Chang, V. Vuletić, and M. D. Lukin, 
Nature Photonics \textbf{8}, 685 (2014).}
\bibitem{Ima}{A. Imamo\u{g}lu, H. Schmidt, G. Woods, and M. Deutsch,
 Phys. Rev. Lett. {\bf 79}, 1467 (1997).}
\bibitem{Birn}{K. M. Birnbaum \textit{et.al.}, 
Nature (London) {\bf 436}, 87 (2005).}
\bibitem{Ada}{A. Miranowicz, M. Paprzycka,Y. X. Liu, J. Bajer, F. Nori, 
Phys. Rev. A {\bf 87}, 023809 (2013).}
\bibitem{Hen}{M. Hennrich, A. Kuhn, and G. Rempe, 
Phys. Rev. Lett. \textbf{94}, 053604 (2005).}
\bibitem{Sch}{S. Schmidt, D. Gerace, A. A. Houck, G. Blatter, H. E. Tureci, 
Phys. Rev. B {\bf 82}, 100507 (2010).}
\bibitem{Sara}{S. Ferretti, L. C. Andreani, H. E. Tureci, D. Gerace, Phys Rev A {\bf 82}, 
013841 (2010).}
\bibitem{Seg}{M. Segev, Y. Silberberg, and D. N. Christodoulides, 
Nature Photonics {\bf 7}, 197 (2013).}
\bibitem{Hart}{M. J. Hartmann, F. G. S. L. Brandão, and M. B. Plenio,
 Laser Photonics Rev. {\bf 2}, 527 (2008).}
\bibitem{Hart2}{M. J. Hartmann, J. Opt. \textbf{18}, 104005 (2016).}
\bibitem{Bran}{B. M. Anderson, R. Ma, C. Owens, D. I. Schuster, and 
J. Simon, Phys. Rev. X \textbf{6}, 041043 (2016).}
\bibitem{atiyah}{M. F. Atiyah , Duality in Mathematics and Physics, 
Lecture notes from the Institut de Matematica de la Universitat de 
Barcelona (2007) (Unpublished).}
\bibitem{Dor}{U. Dorner, R. Demkowicz-Dobrzanski, B. J. Smith, J. S. 
Lundeen, W. Wasilewski, K. Banaszek, and I. A. Walmsley, Phys. Rev. 
Lett, \textbf{102}, 040403 (2009).}
\bibitem{Keb}{K. Jiang, C. J. Brignac, Y. Weng, M. B. Kim, H. Lee, and
 J. P. Dowling, Phys. Rev. A, \textbf{86}, 013826 (2012).}
\bibitem{Sean}{S. D. Huver, C. F. Wildfeuer and J. P. Dowling, Phys. 
Rev. A, \textbf{78}, 063828 (2008).}
\bibitem{Marcel}{M. Bergmann and P. van Loock, Phys. Rev. A \textbf{94},
 012311 (2016).}
\bibitem{bose}{M. H. Yung and S. Bose,Phys. Rev. A, \textbf{71}, 
032310 (2005).}
\bibitem{Kim}{M. S. Kim, W. Son, V. Buzek, and P. L. Knight, Phys. Rev. A 
\textbf{65}, 032323 (2002).}
\bibitem{Chris}{M. Christandl, N. Datta, A. Ekert, and A. J. Landahl, 
Phys. Rev. Lett. \textbf{92}, 187902 (2004).}
\bibitem{Yogesh}{Y. N. Yogleker, C. Thompson, D. D. Scott and G. Vemuri, Eur. 
Phys. J. Appl. Phys, \textbf{63}, 30001 (2013).}
\bibitem{Chris2}{M. Christandl, N. Datta, T. C. Dorlas, A. Ekert, A. Kay, and 
A. J. Landahl, Phys. Rev. A \textbf{71}, 032312  (2005).}
\bibitem{Calic}{M. Calic, P. Gallo, M. Felici, K. A. Atlasov, B. Dwir, 
A. Rudra, G. Biasiol, L. Sorba, G. Tarel, V. Savona, and E. Kapon, 
Phys. Rev. Lett. \textbf{106}, 227402 (2011).}
 \bibitem{Hugg}{A. Huggenberger, S. Heckelmann, C. Schneider,
  S. Höfling, S. Reitzenstein, L. Worschech, M. Kamp and A. Forchel, 
  Appl. Phys. Lett. \textbf{98}, 131104 (2011).} 
 \bibitem{Surr}{A. Surrente, M. Felici, P. Gallo, B. Dwir, A. Rudra, G. Biasiol, 
 L. Sorba, and E. Kapon, Nanotechnology \textbf{22}, 465203 (2011).} 
\end{thebibliography}
\end{document}